\documentclass[letterpaper, 10 pt, conference]{ieeeconf}  

\IEEEoverridecommandlockouts                              

\usepackage{cite}
\usepackage{algorithm}
\usepackage{graphicx}
\graphicspath{ {./figures/} }
\usepackage{xcolor}
\usepackage{hhline}
\usepackage{url}

\usepackage{algpseudocode}
\usepackage{amsmath,amssymb,amsfonts}

\title{\LARGE \bf
Multi-Modal Spectral Parametrization Method (MMSPM) for analyzing EEG activity with distinct scaling regimes
}

\author{Frigyes Samuel Racz$^{1}$, John Milton$^{1}$, Juan Luis Cabrera$^{2}$, Gabor Csukly$^{3}$ and Jose del R. Millan$^{1,4}$
\thanks{$^{1}$Department of Neurology, The University of Texas at Austin, Austin, TX, USA. Email:
        {\tt\small fsr324@austin.utexas.edu}, {\tt\small  john.milton@austin.utexas.edu}}%
\thanks{$^{2}$Departamento de F\'isica Aplicada, Universidad Polit\'ecnica de Madrid, Madrid, Spain. Email:
        {\tt\small juluisca@gmail.com}}%
\thanks{$^{3}$Department of Psychiatry and Psychotherapy, Semmelweis University, Budapest, Hungary. E-mail:
        {\tt\small csukly.gabor@semmelweis.hu}}%
\thanks{$^{4}$Chandra Family Department of Electrical and Computer
Engineering, The University of Texas at Austin, Austin, TX, USA. E-mail:
        {\tt\small jose.millan@austin.utexas.edu}}%
}

\begin{document}

\maketitle
\thispagestyle{empty}
\pagestyle{empty}

\begin{abstract}

Aperiodic neural activity has been the subject of intense research interest lately as it could reflect on the cortical excitation/inhibition ratio, which is suspected to be affected in numerous clinical conditions. This phenomenon is characterized via the aperiodic scaling exponent $\beta$ , equal to the spectral slope following log-log transformation of power spectra. Despite recent progress, however, most current methods do not take into consideration the plausible multimodal nature in the power spectra of neurophysiological recordings - i.e., $\beta$ might be different in low- ($\beta_{lo}$) and high-frequency ($\beta_{hi}$) regimes -, especially in case of $|\beta_{lo}|>|\beta_{hi}|$. Here we propose an algorithm, the multi-modal spectral parametrization method (MMSPM) that aims to account for this issue. MMSPM estimates $\beta_{lo}$ and $\beta_{hi}$ separately using a constrained, piece-wise regression technique, and also assesses if they are significantly different or instead the spectrum is indeed unimodal and can be characterized simply with broadband $\beta$. Here we present the MMSPM algorithm and evaluate its performance \textit{in silico} on simulated power spectra. Then, we use MMSPM on resting-state electroencephalography (EEG) data collected from 19 young, healthy volunteers, as well as on a separate dataset of EEG recordings from 30 schizophrenia patients and 31 healthy controls, and demonstrate that broadband (0.1-100 Hz and 0.5-45 Hz) EEG spectra can indeed present a bimodality pattern with significantly steeper low-range ($<\sim2$ Hz) and flatter high-range scaling regimes (i.e., $|\beta_{lo}|>|\beta_{hi}|$).


\indent \textit{Clinical relevance} — The MMSPM method characterizes aperiodic neural activity in distinct scaling regimes, which can be relevant in numerous pathological conditions such as dementia or schizophrenia.
\end{abstract}

\section*{DISCLOSURE}

$\copyright$ 2025 IEEE. Personal use of this material is permitted. Permission from IEEE must be obtained for all other uses, in any current or future media, including reprinting/republishing this material for advertising or promotional purposes, creating new collective works, for resale or redistribution to servers or lists, or reuse of any copyrighted component of this work in other works.

\section{INTRODUCTION}

It is widely acknowledged that neural activity is composed of rhythmic patterns with characteristic frequencies intermixed with arrhythmic activity \cite{he2010temporal}. In the power spectrum of electrophysiological neural recordings this phenomenon manifests as narrow-band peaks being superimposed on the broadband, aperiodic 'background' component \cite{he2010temporal}. Since the generating mechanisms of aperiodic and oscillatory brain activity are likely different \cite{buzsaki2012origin}, it is important to treat them separately as changes in the aperiodic component can confound oscillatory power estimates and vice versa \cite{wen2016separating,donoghue2020parameterizing}. For example, some observed alterations in scale-free electroencephalography (EEG) power spectra in schizophrenia can be attributed to changes in the scale-free, but not narrow-band oscillatory activity \cite{racz2021separating,racz2025reduced}.

The broadband component of EEG spectra is often considered scale-free (or \textit{fractal}) as $Power\propto f^{-\beta}$ and characterized via its scaling exponent $\beta$ \cite{eke2002fractal,he2014scale}. Since the power-law relationship reduces to a linear one following log-transformation, $\beta$ is also often referred to as the spectral slope. Treating aperiodic and oscillatory brain activity separately is impediment: an observed increase in spectral power in a narrow  frequency band (e.g., 8-12 Hz for alpha activity) in itself might reflect increased alpha synchronization, but can also result from aperiodic activity becoming steeper, potentially leading to misinterpretations \cite{donoghue2020parameterizing}. On the other hand, the presence of oscillatory peaks can substantially bias spectral slope estimates \cite{he2010temporal,wen2016separating}. In line, the current emphasis has been directed towards developing computational tools capable of separating the broadband from oscillatory components in the power spectrum \cite{wen2016separating}, with the recently introduced 'fitting oscillations \& one over f' (FOOOF) method gaining immense utility recently \cite{donoghue2020parameterizing}.

While FOOOF is an excellent algorithm and invaluable tool for neuroscientists, a potential limitation is that it assumes a Lorentzian structure of the aperiodic activity \cite{donoghue2020parameterizing}. The consequence of this assumption is that FOOOF only considers a single scaling range with a potential 'knee' in the low-frequency end. However, it is frequently observed that neural activity can express multimodality, in that different scaling ranges might be characterized with different scaling exponents/slopes, most commonly with a flatter low-range and steeper high-range regime \cite{miller2009power,he2010temporal,nagy2017decomposing,racz2021separating,racz2025reduced}. Recent preliminary evidence also suggests that the bimodality pattern of low- ($\beta_{lo}$) and high- ($\beta_{hi}$)-range exponents ($\beta_{lo}<\beta_{hi}$) might be eliminated or even reversed by certain, common data manipulations such as spatial filtering \cite{racz2024alternative}. Furthermore, while it appears established that a decrease in cortical excitation/inhibition (E/I)-ratio is followed by a steepening of the power spectrum (increase in $\beta$) above $\sim20$ Hz \cite{gao2017inferring}, it is also indicated that this pattern might be the opposite in the low-frequency (below $\sim5$ Hz) regime \cite{becker2018alpha,muthukumaraswamy20181}. 

These observations suggest that techniques that can automatically account for multimodality in the broadband $1/f$ component, regardless of $\beta_{lo}<\beta_{hi}$ or the opposite, could be a valuable tool for a more wholesome characterization of neural power spectra. Here we present such a tool, which we term multi-modal spectral parametrization method (MMSPM). First we provide the outline of the algorithm, then evaluate it first \textit{in silico} on simulated power spectra, then on real electroencephalography (EEG) recordings collected from healthy volunteers.

\begin{figure*}[!t]
    \centering
    \includegraphics[width=\textwidth]{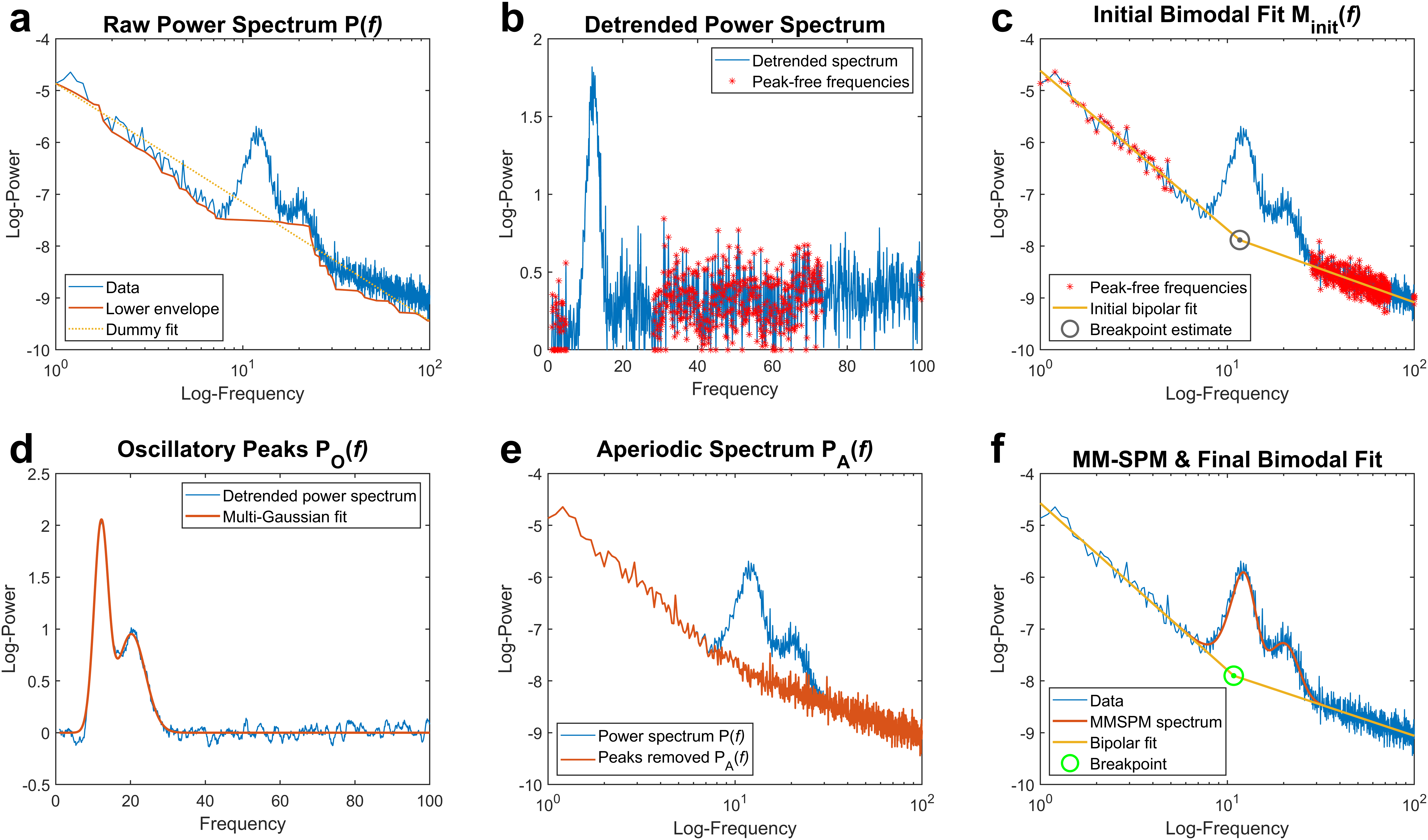}
    \caption{Steps of the MMSPM procedure. Power spectrum was simulated with $\beta_{lo}=1.5$ and $\beta_{hi}=0.5$ with $f_{c}$ at 10 Hz, at a noise-to-signal ratio of 0.15. Two oscillatory peaks were added at 12 and 20 Hz with relative peak amplitudes 2:1 and $\sigma$ 2:4, emulating neural activity.}
    \label{fig:MMSPM_steps}
\end{figure*}

\section{METHODS}

\subsection{The MMSPM algorithm}

Similar to previous, time-domain techniques for analyzing scale-free processes \cite{nagy2017decomposing}, the main purpose of MMSPM is to account for the possibility that the broadband, aperiodic component consists of two (low and high) scaling regimes with distinct scaling exponents $\beta_{lo}$ and $\beta_{hi}$, respectively. The two regimes are assumed to be separated at a singular frequency $f_{bp}$, producing a \textit{breakpoint} in the spectrum $P(f)$ when visualized on a log-log scale. While FOOOF models the aperiodic component $P_{A}(f)$ with a Lorentzian as $log(P_{A}(f))=b-log(k+f^{-\beta})$ with $b$ as the log-offset and $k$ as the 'knee' parameter allowing for a convex bend in the low-frequency end, the main feature of MMSPM is that assumes a piece-wise power-law model such as

\begin{equation}
    \begin{split}
    P_{A}(f)=
        \begin{cases}
        c_{1}\times f^{-\beta_{lo}} | f<=f_{bp}, \\
        c_{2}\times f^{-\beta_{hi}} | f<=f_{bp} \\
        \end{cases}
        \\
        \text{with } c_{1}\times f_{bp}^{-\beta_{lo}}=c_{2}\times f_{bp}^{-\beta_{hi}}
        \label{eq:MMSPM_model}
        \end{split}
    \end{equation}

\noindent with imposing the constraint of assuming continuity in the aperiodic component. As an algorithm, MMSPM builds strongly on the foundations of FOOOF \cite{donoghue2020parameterizing} with some key modifications; in what follows we describe each step of the MMSPM procedure, highlighting the differences from FOOOF. The current Matlab implementation of MMSPM is available in the GitHub repository at \url{https://github.com/samuelracz/MMSPM} (\textit{in development}). The steps of MMSPM are illustrated on \textbf{Figure \ref{fig:MMSPM_steps}}.

MMSMP takes as input spectral power obtained e.g., via Welch's periodogram method, the corresponding frequencies, and a set of model parameters (detailed below). As a first step, power values are log-transformed, and all subsequent operations are performed on log-transformed data unless stated otherwise (\textbf{Figure \ref{fig:MMSPM_steps}a}). Next, an initial model guess is obtained. In that, the spectrum is first detrended approximately using a simple power-law fit, then local maxima (i.e., peaks) are detected by a peak-detection algorithm ({\fontfamily{qcr}\selectfont{findpeaks.m}} in Matlab). The goal is to isolate those segments in the spectrum that are absent of peaks (\textbf{Figure \ref{fig:MMSPM_steps}b}). Note that in case no peaks are detected automatically, instead $P(f)$ is detrended using a lower envelope obtained via a rolling minimum (assuming a power-law decay in $P(f)$) and frequencies with power in a pre-defined lower quantile are selected, likely not corresponding to peaks \cite{donoghue2020parameterizing}. Once this is achieved, an initial model $M_{init}(f)$ is obtained according to Eq. (\ref{eq:MMSPM_model}) using only these subset of frequencies (\textbf{Figure \ref{fig:MMSPM_steps}c}). Importantly, the high-frequency regime is usually overrepresented compared to the low-frequency regime due to the power-law decay \cite{eke2002fractal,wen2016separating}. To counter this, in the cost function \textit{i}) error terms (residuals) are weighted according to $f^{-\omega}$ to penalize errors in the low-range more, and \textit{ii}) the mean squared error (MSE) of residuals is computed separately for low- and high-range and then summed. $M_{init}(f)$ is then removed from the power spectrum to isolate the component containing oscillatory peaks without background activity ($P_{O}(f)$). In our evaluations, we have found $\omega=0.25$ as a suitable value that works well with most simulated and real-world data.

In a similar vein to FOOOF, oscillations $G_{n}(f)$ are modeled via Gaussians as 

\begin{equation}
    G_{n}(f)=A\times e^{-\frac{1}{2}(\frac{f-f_{c}}{\sigma})^{2}}
    \label{eq:gaussian}
\end{equation}

\noindent in an iterative manner (see \cite{donoghue2020parameterizing} for details), with $A$, $f_{c}$ and $\sigma$ being the log-amplitude, center frequency and standard deviation, respectively. In every iteration $n$, a simple, most prominent Gaussian $G_{n}(f)$ is detected in $PO(f)$. If the amplitude of $G_{n}(f)$ surpasses a peak threshold (by default 2 times the current $\sigma$ of $P_{O}(f)$) then $G_{n}(f)$ is removed and the procedure is repeated until no more supra-threshold peaks are detected or the number of maximal allowed peaks (set in model settings) is reached. Out of the detected Gaussians, those that overlap too much (i.e., $|f_{c,i}-f_{c,j}|<0.25\times min(\sigma_{i},\sigma_{j})$) and those too close to the edge of the full broadband range ($f_{c}$ less than $1\times \sigma$ from edge) are dropped (for overlapping $G_{n}(f)$ the one with smaller $A$). Then, the oscillatory component is modeled collectively as a sum of Gaussians 

\begin{equation}
    P_{O}(f)=\sum_{n=1}^{N}G_{n}(f)
    \label{eq:mixed_gaussian}
\end{equation}

\noindent with $N$ denoting the remaining number of peaks, with the previously obtained individual peak parameters ($A_{i}$,$f_{c,i}$ and $\sigma_{i}$) used as initial guesses when fitting the model for $P_{O}(f)$ (\textbf{Figure \ref{fig:MMSPM_steps}d}). $P_{O}(f)$ is then removed from $P(f)$ to obtain $P_{A}(f)$ (\textbf{Figure \ref{fig:MMSPM_steps}e}). Finally, from $P_{A}(f)$ the piece-wise power-law model $M(f)$ in the form of Eq. (\ref{eq:MMSPM_model}) is obtained in a similar manner as $M_{init}(f)$ (\textbf{Figure \ref{fig:MMSPM_steps}f}). In summary, MMSPM provides the following parametrization of the power spectrum $P(f)$: for $P_{A}(f)$, it yields $\beta_{lo}$, $\beta_{hi}$, their corresponding offsets $c_{lo}$ and $c_{hi}$, and the estimated breakpoint frequency $f_{bp}$. For $P_{O}(f)$, it returns $A_{n}$, $f_{c,n}$ and $\sigma_{n}$ for each oscillation $G_{n}(f)$.

It is important to highlight that MMSPM will enforce a bimodal model even when the spectrum might be unimodal i.e., $P_{A}(f)$ is characterized with a single $\beta$. To test this, a test statistic $t_{MMSPM}$ is constructed to test if $\beta_{lo}$ and $\beta_{hi}$ estimates are significantly different, as

\begin{equation}
    t_{MMSPM}=\frac{\beta_{lo}-\beta_{hi}}{\sqrt{SE_{lo}^{2}+SE_{hi}^{2}}}
    \label{eq:test_statistic}
\end{equation}

\noindent where $SE_{lo}$ and $SE_{hi}$ are the corresponding standard errors of linear regression models of log-power on log-frequency with parameters obtained from $M(f)$. Then, $t_{MMSPM}$ is contrasted against a t-distribution with degrees of freedom $df=min(Nf_{lo},Nf_{hi})-2$, where $Nf_{lo}$ and $Nf_{hi}$ equal to the number of frequencies in the low- and high-regimes, respectively. If $\beta_{lo}$ and $\beta_{hi}$ are not statistically significantly different at $p<0.05$, instead a simple power-law model of $P_{A}(f)=c\times f^{-\beta}$ is obtained.

\subsection{In silico evaluation on simulated power spectra}

MMSPM was first tested on simulated power spectra. In that, power spectra were simulated according to the model

\begin{equation}
    P(f)=P_{A}(f)+\sum_{n=1}^{N}G_{n}(f)+\epsilon(f)
\end{equation}

\noindent with $\epsilon(f)$ being added uniform noise. Iterations of $P(f)$ were generated in the frequency range $[0.1:100]$ Hz at 0.1 Hz resolution, to allow for four orders of magnitude in the log-frequency range. $\beta_{lo}$ was varied between $[0.5:1.5]$ and $\beta_{hi}$ between $[0.5:2.5]$ allowing for both convex ($\beta_{lo}<\beta_{hi}$) and concave ($\beta_{lo}>\beta_{hi}$) cases. Breakpoint position $f_{bp}$ was varied between $[5:15]$ Hz. In these preliminary simulations only a single peak was introduced, whose $f_{c}$ varied between $[10:14]$ Hz, $\sigma$ between $[1:2]$ Hz and $A$ between $[0.5:3.5]$ AU. For each parameter combination (35 in total), $N=1000$ realizations were executed. Model performance was captured in the MSE of $P_{A}(f)$ and $P_{O}(f)$ parameter estimates in contrast to ground truth values, instead of goodness-of-fit measures of $M(f)$ itself \cite{donoghue2020parameterizing}.

\begin{figure*}[!t]
    \centering
    \includegraphics[width=\textwidth]{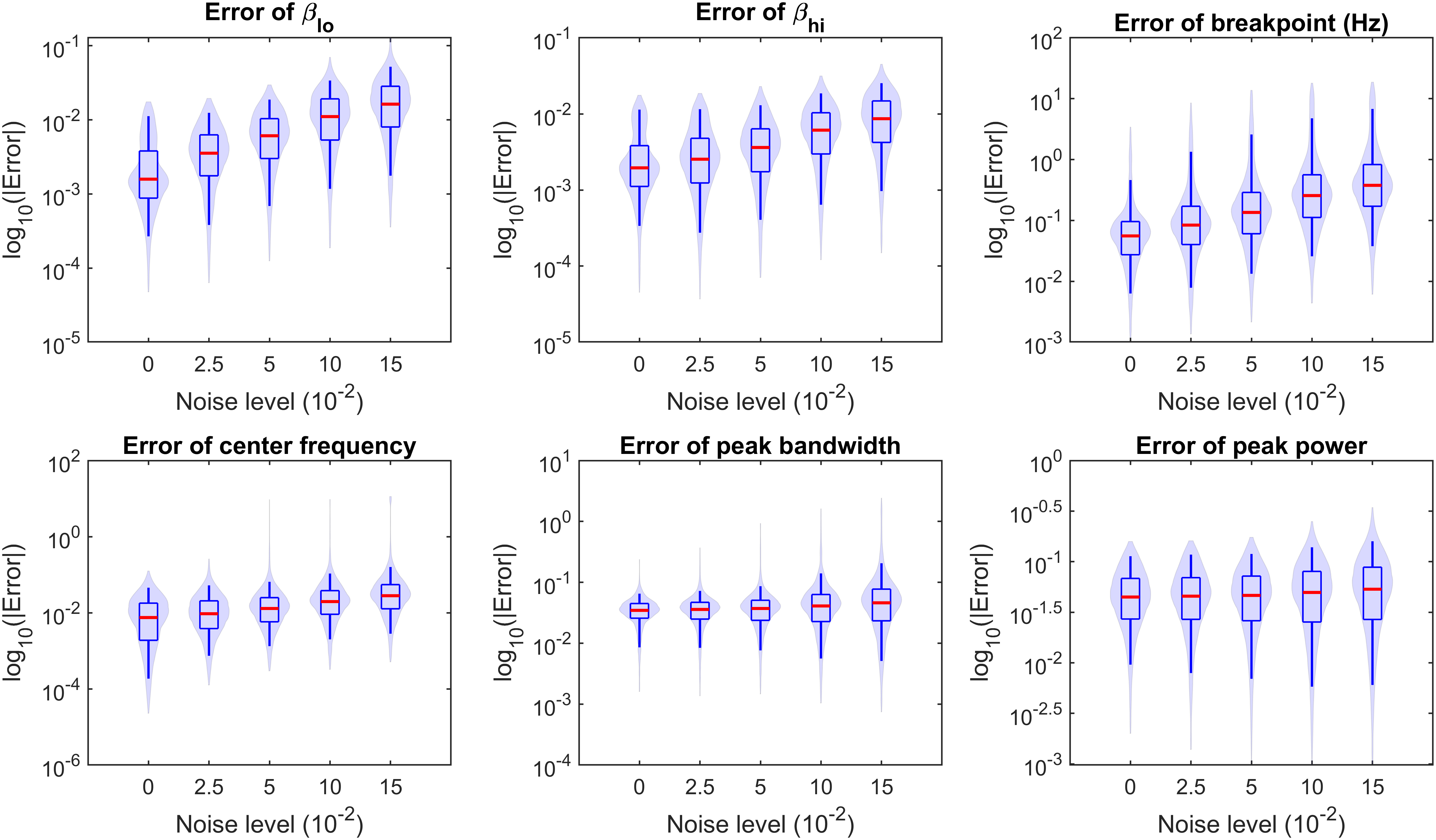}
    \caption{MMSPM performance evaluation. Log-transformed absolute error in endpoint parameters are shown at variuos levels of noise.}
    \label{fig:MMSPM_error}
\end{figure*}

\subsection{Testing on EEG data}

We also tested MMSPM on two independent EEG datasets. In the first dataset, EEG data was collected from $N=19$ young, healthy volunteers (mean age $27.26\pm 6.11$ years, 10 females, all right handed). EEG measurements were carried out in eyes-open (EO) and eyes-closed (EC) resting-states ($\sim2$ minutes each) using a 32-channel ANT Neuro amplifier at 512 Hz sampling (Ref: Iz, Ground: AFz, for an illustration of the electrode montage, see Figure 4 in \cite{kumar2023cognidavinci}). From both states, 110-seconds of continuous data was selected, bandpass-filtered with a $4^{th}$-order zero-phase Butterworth filter with cutoff frequencies 0.025 and 164 Hz, then transformed to reference-free current source density (CSD) \cite{perrin1987scalp} time series. Power spectra were obtained between 0.1 and 100 Hz at a resolution of 0.1 Hz using Whelch's periodogram method with a window size of 20 seconds, window step size of 5 seconds, and Hanning windowing. Finally, spectra from all channels were averaged to obtain a global estimate. MMSPM was then used to evaluate EO with EC spectra in the study group. The study was reviewed and approved by UT Austin's Institutional Review Board (approval number: 2020-03-0073).

We also utilized MMSPM to analyze an openly available dataset \cite{racz2025dataset} containing EC resting-state EEG data obtained from 30 patients with schizophrenia (SZ) and 31 age- and sex-matched healthy controls (HC). Data was collected from 55 cortical locations at 256 Hz sampling with a Neuroscan amplifier. The Semmelweis University Regional and Institutional Committee of Science and Research Ethics reviewed and approved the study (approval number: 197/2015). For EEG pre-processing and study cohort detail, please see \textbf{Table 1} in \cite{racz2025reduced}. Here, global average spectra obtained from CSD-transformed, 30-second EEG data were analyzed in the 0.5-45 Hz regime.

In both protocols, all participants provided written informed consent prior to the EEG measurement, and both studies were conducted in line with the Declaration of Helsinki.

\section{Results and Discussion}

\textbf{Figure \ref{fig:MMSPM_error}} shows the outcomes of the \textit{in silico} evaluation. The median absolute error for both $\beta_{lo}$ and $\beta_{hi}$ was below 0.1 even at more excessive levels of noise (0.15 noise-to-signal ratio), while the breakpoint was detected with a $\pm1$ Hz precision on average. The center frequency $f_{c}$ and bandwidth $\sigma$ was detected with similar precision, and peak power estimates were found about 0.1 AU for most noise levels. Performance in estimating spectral parameters are in the same range as those of previous methods in characterizing aperiodic and oscillatory properties \cite{wen2016separating,donoghue2020parameterizing}. While here we only report performance evaluation in the presence of a singular oscillatory peak, similar outcomes were found when evaluating MMSPM on spectra with 2 or 3 peaks (see e.g., \textbf{Figure \ref{fig:MMSPM_steps}}). 

\begin{figure}[!t]
    \centering
    \includegraphics[width=250pt]{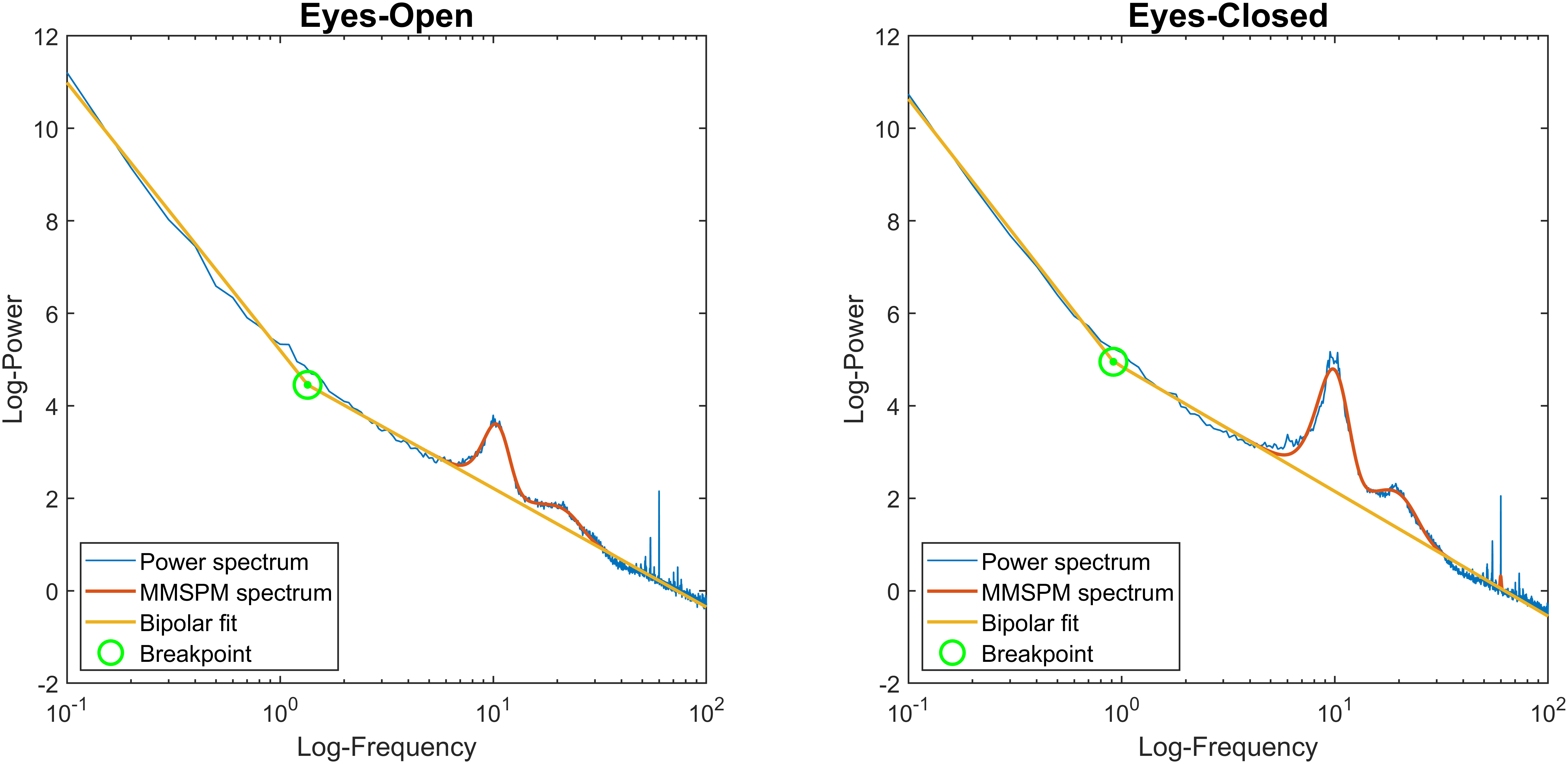}
    \caption{Grand average power spectra in eyes-open (EO, left) and eyes-closed (EC, right). On the group level, significant bimodality with steeper low-frequency regime is observed with a breakpoint at 1.35 Hz in EO and 0.92 Hz in EC. While alpha (EO: 10.39 Hz, EC: 9.87 Hz) and beta (EO: 21.16 Hz, EC: 19.62 Hz) are observed in both states, a clear increase in both frequency bands is observable when transitioning from EO to EC.}
    \label{fig:EO_vs_EC}
\end{figure}

\textbf{Figure \ref{fig:EO_vs_EC}} illustrates the performance of MMSPM on real EEG data. In both EO (left) and EC (right) states a concave bimodality is apparent, with a steeper low-, and a flatter high-frequency regime. This bimodality was confirmed as significant by MMSPM for 17 and 18 out of 19 participants for EO and EC, respectively. On the group level, bimodality was tested with a two-way repeated measures ANOVA including {\fontfamily{qcr}\selectfont{state}} (EO vs. EC) and {\fontfamily{qcr}\selectfont{range}} (low- vs. high-frequency) fixed effects (after confirming assumptions of normality via Lilliefors test). This model indicated a significant main effect of {\fontfamily{qcr}\selectfont{range}} ($df=18$, $F=141.95$, $p<10^{-9}$). Post-hoc pairwise comparisons revealed $\beta_{lo}>\beta_{hi}$ for both EO ($2.3814\pm 0.6805$ vs. $1.0313\pm 0.1314$, paired t-test, $p<10^{-6}$, $t_{18}=9.2757$) and EC ($2.6318\pm 0.6948$ vs. $1.0663\pm 0.1771$, paired t-test, $p<10^{-6}$, $t_{18}=9.1344$). There was no significant {\fontfamily{qcr}\selectfont{state}} main ($p=0.2196$) or {\fontfamily{qcr}\selectfont{state}} $\times$ {\fontfamily{qcr}\selectfont{range}} interaction effect ($p=0.3025$), indicating that closing the eyes had no effect on global spectral slopes. There was a significant EO-to-EC increase in isolated alpha peak power ($1.2664\pm 0.9450$ vs. $2.6245\pm 0.8939$, paired t-test, $p<10^{-5}$, $t_{18}=6.8126$) as well as beta peak power ($0.5832\pm 0.4976$ vs. $1.0694\pm 0.5691$, paired t-test, $p<0.0002$, $t_{18}=4.7064$), in line with expectations \cite{barry2007eeg}. While it appears that the breakpoint shifted towards the lower frequencies when opening the eyes, this tendency did not reach the level of statistical significance.

\begin{figure}[!t]
    \centering
    \includegraphics[width=250pt]{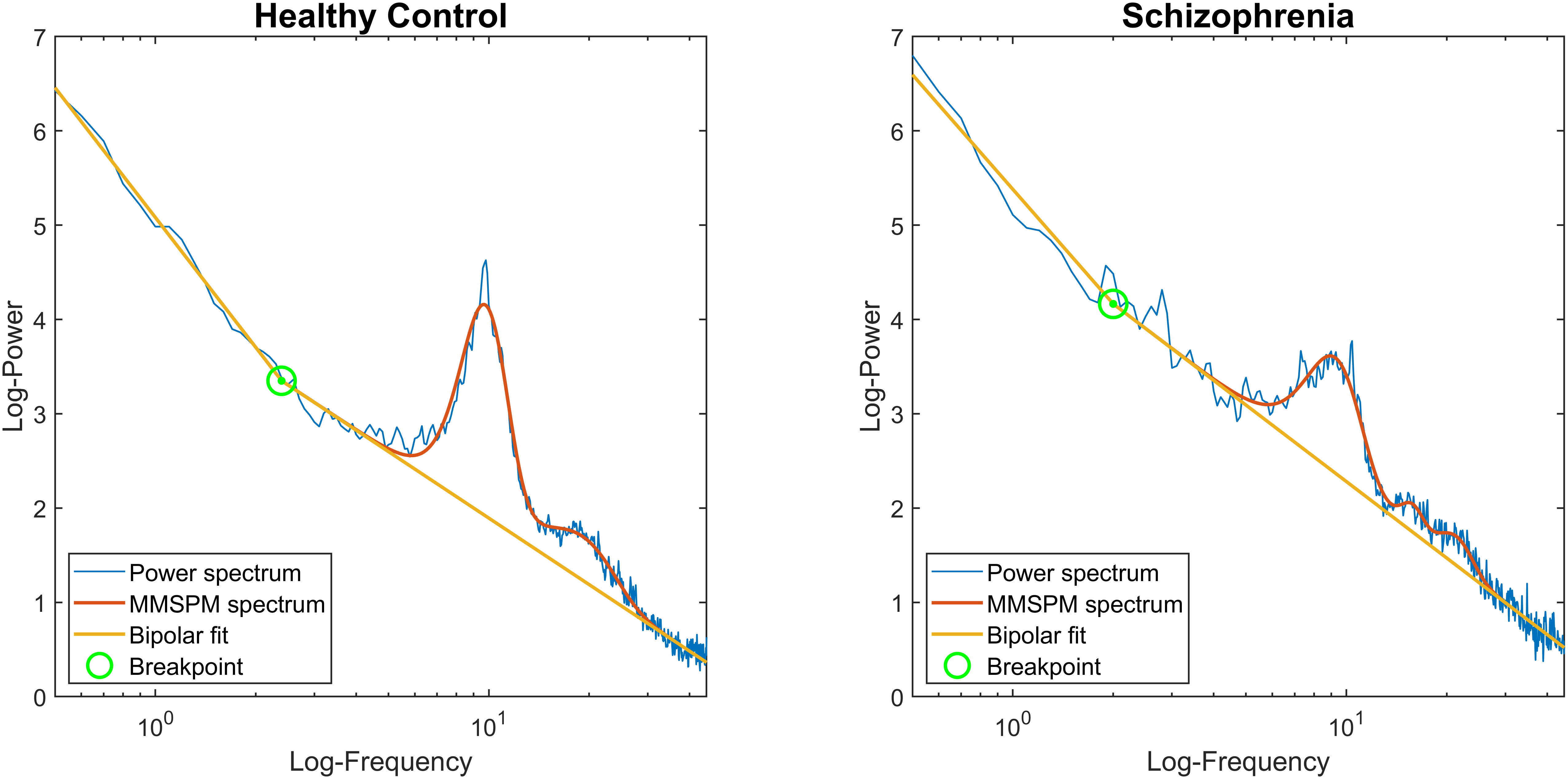}
    \caption{Grand average power spectra in healthy controls (HC, left) and schizophrenia patients (SZ, right). On the group level, significant bimodality with steeper low-frequency regime is observed with a breakpoint at 2.39 Hz in HC and 2.0 Hz in SZ. While a singular beta peak is observed in HC, two distinct peaks are visible in the beta band in SZ.}
    \label{fig:HC_vs_SZ}
\end{figure}

\textbf{Figure \ref{fig:HC_vs_SZ}} shows grand average spectra from the HC (left) and SZ (right) groups. Concave bimodality is apparent yet again, with $\beta_{lo}>\beta_{hi}$. Significant bimodality was only confirmed 10 and 8 participants from the HC and SZ groups, respectively, likely due to the smaller scaling range and consequently more substantial standard error. However, group-level analysis still indicated significant bimodality within both HC (1.5223, IQR: $[1.2302; 1.7024]$ vs. 1.0367, IQR: $[0.7328; 1.1194]$, Wilcoxon signed rank t-test, $p=0.0018$, $z=3.1159$) and SZ ($1.4018\pm 0.5128$ vs. $0.9687\pm 0.3488$, paired t-test, $p=0.0013$, $t_{29}=3.5515$) groups. No HC vs. SZ between-group difference was found in either $\beta_{lo}$ or $\beta_{hi}$, in line with our previous findings \cite{racz2025reduced}. Interestingly, instead of the previously observed $\beta_{lo}<\beta_{hi}$-type bimodality when fitting spectral slopes on the 1-4 and 20-45 Hz regimes independently \cite{racz2025reduced}, here with MMSPM we see this pattern reversed, with $f_{bp}$ close to 2 Hz. This for one highlights the importance of the continuity constraint introduced in Eq. (\ref{eq:MMSPM_model}); however, bimodality is also likely influenced by applying a different spatial filtering scheme, which has been shown to have such effects on bimodal slope estimates \cite{racz2024alternative}. In addition, significant between-group difference was observed in isolated alpha peak power, with lower values in the SZ group (HC: $1.9871\pm 0.9548$ vs. SZ: $1.4213\pm 0.7503$, independent two-sample t-test, $p=0.0128$, $t_{59}=2.5677$). While it would require more elaborate analysis to infer on the pathophysiological relevance of these EEG patterns (considering e.g., disease phenotype and medication \cite{racz2025reduced}) that is beyond the scope of this report, these results illustrate the utility of MMSPM on EEG data collected from clinical populations.

Findings reported here indicate that a bimodal approach to aperiodic neural activity is recommended. In particular, it is relevant to utilize a model that is able to account for a bimodality pattern where $\beta_{lo}>\beta_{hi}$, as we have found in both our cohorts regarding global EEG activity. In this regard MMSPM overcomes the limitation of FOOOF of only being able to treat cases of convex bimodailty in the aperiodic component (only demonstrated via \textit{in silico} data here), while providing an equivalent parametrization of oscillatory components. Similar approaches have been introduced previously to treat bimodality in the time domain, yet those focused on the fractal scaling exponents without eliminating the potential bias from oscillatory signal components \cite{nagy2017decomposing,mukli2018impact}. Therefore, we believe MMSPM is indeed filling a knowledge gap and can be a valuable tool for neuroscientists. In line with that, we have the source code openly available (see \textbf{Methods}) and intend on further developing this method in the future.

The biological basis of the observed $\beta_{lo}>\beta_{hi}$-type bimodality is yet to be understood. Similar bimodality pattern can be observed in case of the superposition of two, independent scale-free processes \cite{echeverria2016linear}, indicating different generating mechanisms for low- and higher-frequency brain activity. It is believed that very-low frequency (i.e., $<1.5Hz$, VLF) EEG oscillations represent the activity of large-scale brain networks and might be clinically relevant in pathologies such as attention deficit/hyperactivity disorder \cite{helps2010altered} or sleep disorders \cite{g2022overnight}. MMSPM can serve as a useful tool for similar studies to characterize the spectral slope in the VLF regime, avoiding potential bias from higher-frequency activity with smaller $\beta_{hi}$. While in the analyzed frequency range we did not observe the more common $\beta_{lo}<\beta_{hi}$ bimodality pattern \cite{miller2009power,he2010temporal}, our simulations indicate that MMSPM can be an equally valuable tool in such cases as well.

It is important to disclose that MMSPM is still under development and here we only report outcomes obtained in the bimodal case. However, previous analysis of electrocorticography data indicates that the broadband spectrum of neural activity might be composed of more that two, distinct scaling regimes \cite{he2010temporal}. One of our goals in the future is to extend the current MMSPM model to be able to handle cases with an arbitrary number of scaling regimes. Also, not surprisingly, simulation results indicate that model performance deteriorates in the presence of noise. This stresses the fact that for reliable performance, one needs to obtain high-quality estimates of the power spectrum. In this proof-of-concept analysis we opted for analyzing global power spectra obtained from relatively long, continuous EEG recordings so to ensure that power spectral estimates were robust and noise free. A more in-depth evaluation of MMSPM - beyond the scope of the current report - is required to assess its performance on power spectra obtained from shorter (e.g., 4-second) EEG epochs or less broad frequency regimes (e.g., 1-45 Hz). In particular, the influence of the absolute (i.e., in Hz) and relative (i.e., at $\%$ of the considered frequency regime) position of $f_{bp}$ needs to be systematically investigated. Here we limited ourselves to varying $f_{bp}$ between $[5:15]$ Hz, empirical data shows that $f_{bp}$ can be positioned outside this range. Based on these data sets, MMSPM could handle these cases well; nevertheless, this needs to be addressed rigorously as well in the future.

\section{Conclusions}

Evaluation of both simulated and real EEG power spectra indicates that MMSPM can be a useful and effective tool in characterizing neural activity, filling an important gap by being able to account for both convex- and concave-type bimodality of aperiodic neural activity. Future work should focus on assessing its performance and utility in more empirical settings, with a special focus on pathological conditions where aperiodic neural activity appears to be affected. In particular, it would be interesting to evaluate the utility of MMSPM for E/I ratio assessment in cases where excitability/inhibitory tone is experimentally modulated e.g., via transcranial electrical \cite{lefaucheur2019mechanisms} or magnetic \cite{fitzgerald2006comprehensive} stimulation. Relatedly, MMSPM can also serve as a valuable tool in the investigation of clinical conditions --- besides schizophrenia \cite{lanyi2024excitation} --- where the E/I ratio is a suspected pathological factor, such as Alzheimer's Disease \cite{martinez2023combining} or autism spectrum disorder \cite{bruining2020measurement}.

\section*{ACKNOWLEDGMENT}

This work was partially funded by the Charley Sinclair foundation and the Coleman Fung foundation.

\nocite{*}
\bibliographystyle{unsrt}
\bibliography{ref}

\end{document}